\documentclass[aps,prd,twocolumn,notitlepage,nofootinbib,floatfix,superscriptaddress]{revtex4-1}
\usepackage{amsmath, amsthm, amssymb, amsfonts, amsbsy,mathrsfs}

\usepackage{calligra,bm}
\usepackage{stmaryrd}

\usepackage[hidelinks,bookmarks=true]{hyperref} 
\hypersetup{pdfstartview=FitH,pdfhighlight=/O,colorlinks=false}

\begin{document}

\title{Some globally conserved currents from generalized Killing vectors and scalar test fields}

\author{Justin C. Feng}
	\affiliation{Theory Group, Physics Department, University of Texas at Austin, Texas 78712, USA}
	\affiliation{St. Edward's University, Austin, Texas 78704, USA}

\preprint{UTTG-12-18}


\begin{abstract}
In this article, I discuss the construction of some globally conserved currents that one can construct in the absence of a Killing vector. One is based on the Komar current, which is constructed from an arbitrary vector field and has an identically vanishing divergence. I obtain some expressions for Komar currents constructed from some generalizations of Killing vectors which may in principle be constructed in a generic spacetime. I then present an explicit example for an outgoing Vaidya spacetime which demonstrates that the resulting Komar currents can yield conserved quantities that behave in a manner expected for the energy contained in the outgoing radiation. Finally, I describe a method for constructing another class of (non-Komar) globally conserved currents using a scalar test field that satisfies an inhomogeneous wave equation, and discuss two examples; the first example may provide a useful framework for examining the arrow of time and its relationship to energy conditions, and the second yields (with appropriate initial conditions) a globally conserved energy- and momentumlike quantity that measures the degree to which a given spacetime deviates from symmetry.
\end{abstract}


\maketitle


\section{The Komar current}
In a 1959 article \cite{Komar1959}, Arthur Komar presented a globally conserved current (the Komar current) for general relativistic spacetimes, which is constructed from an arbitrary vector field $U^\mu$ (which serves as the generator for diffeomorphisms). The Komar current is of the form:\footnote{Note that there is a gauge freedom in this definition; for a torsion-free connection $\nabla_\mu$, $J^\mu_K$ is invariant under the transformation $U^\mu \rightarrow U^\mu +\nabla^\mu \sigma$ ($\sigma$ being a scalar field).}
\begin{equation} \label{CCGR-Komar}
J^\mu_K := \nabla_\nu \left( \nabla^\mu U^\nu - \nabla^\nu U^\mu \right) .
\end{equation}

\noindent The Komar current has theoretical appeal because it can be derived from the action in the context of N{o}ether's theorem---in \cite{Katz1985,Baketal1993,Obukhovetal2006a,*Obukhovetal2006b}, it was shown that the Komar current is in fact the conserved current corresponding to the diffeomorphism invariance of the Einstein-Hilbert Lagrangian coupled to matter.\footnote{I also refer the reader to \cite{KBL1997} and related work which extend the analysis to more general theories of gravity \cite{Deruelleetal2004,Obukhovetal2006c,*Obukhovetal2013,*Obukhovetal2014,*Obukhovetal2015,SchmidtBicak2018}.} 

Using the Ricci identity $[\nabla_\mu,\nabla_\nu] T^{\alpha \beta} = R^\alpha{_{\sigma \mu \nu}} T^{\sigma \beta} + R^\beta{_{\sigma \mu \nu}} T^{\alpha \sigma}$ for the Levi-Civita connection $\nabla_\mu$, it is straightforward to show that the divergence of $J^\mu_K$ identically vanishes:
\begin{equation} \label{CCGR-KomarDivergence}
\begin{aligned}
\nabla_\mu J^\mu_K 
	&= [ \nabla_\mu , \nabla_\nu ] \nabla^\mu U^\nu \\
	&= R{^\mu}{_{\sigma \mu \nu}} \nabla^\sigma U^\nu + R{^\nu}{_{\sigma \mu \nu}} \nabla^\mu U^\sigma \\
	&= R{_{\sigma \nu}} \nabla^\sigma U^\nu - R{_{\sigma \mu}} \nabla^\mu U^\sigma 
	=0 .
\end{aligned}
\end{equation}

\noindent The above result is an identity for any quantity of the form given in Eq. (\ref{CCGR-Komar}); it only depends on the Ricci identity for rank-2 tensors. Note also that Eq. (\ref{CCGR-KomarDivergence}) permits a shift freedom in $J^\mu_K$; any divergence-free vector added to $J^\mu_K$ preserves the divergence-free property Eq. (\ref{CCGR-KomarDivergence}). One potentially useful example is a vector field of the form $\nabla^\mu \phi$, which is divergence free if the scalar field satisfies the wave equation $\Box \phi = 0$.

From the Komar current $J^\mu_K$, I may construct the 3-form:
\begin{equation} \label{CCGR-Komar3Form}
J^\mu_K \, d\Sigma_\mu = 2 \nabla_\nu \left(\nabla^{[\mu} U^{\nu]}\right) d\Sigma_\mu.
\end{equation}

\noindent By Stokes' theorem, the integral of the above over some constant-time hypersurface $\Sigma_t$ becomes:
\begin{equation} \label{CCGR-Komar3Form}
\int_{\Sigma_t} J^\mu_K \, d\Sigma_\mu = 2 \int_{\partial \Sigma_t} \nabla^{[\mu} U^{\nu]}\, dS_{\mu \nu}.
\end{equation}

\noindent where $d\Sigma_\mu$ and $dS_{\mu \nu}$ are the respective surface elements\footnote{See Ch. 3 of \cite{Poisson} for the explicit expressions.} for $\Sigma_t$ and $\partial \Sigma_t$. Equation (\ref{CCGR-Komar3Form}) may be used to construct quasilocal expressions for quantities associated with a Komar current. Note in a closed universe, the constant-time hypersurfaces $\Sigma_t$ are compact and without boundary, so that the Komar integrals vanish identically.

For a Killing vector\footnote{See the Appendix for a discussion of Killing vectors and conservation laws.} $\chi^\mu$ , Eq. (\ref{CCGR-Komar3Form}) is the Komar integral, and it is straightforward to show that the Komar current $J^\mu_K$ [Eq. (\ref{CCGR-Komar})] becomes:
\begin{equation} \label{CCGR-KomarKilling}
\begin{aligned}
J^\mu_\chi 
&= - 2 \, \Box \chi^\mu = 2 \, R^\mu{_\nu} \chi^\nu .
\end{aligned}
\end{equation}

\noindent where I have made use of the following expression, which may be obtained from the divergence of Killing's equation $\nabla_{(\mu} \chi_{\nu)}=0$ [cf Eq. (\ref{CCGRA-KillingWaveEquation})]:
\begin{equation} \label{CCGR-KillingWaveEquation}
\Box \chi^{\mu} + R{^\mu}{_\nu} \, \chi^{\nu} = 0 .
\end{equation}

\noindent Given a Killing vector $\chi^\mu$, Eq. (\ref{CCGR-KomarKilling}) may then be used in conjunction with the Komar integrals Eq. (\ref{CCGR-Komar3Form}) to obtain (up to factors of two) the Komar mass and angular momentum \cite{Komar1962}.


\section{Komar current from almost Killing vectors}
In spacetimes which do not admit Killing vectors, one may consider the construction of Komar currents from some generalization of Killing vectors. The general idea was originally proposed by Komar in \cite{Komar1962}, in which he considered semi-Killing vectors, which are divergence-free and (under the divergence-free condition) satisfy an equation equivalent to Eq. (\ref{CCGR-KillingWaveEquation}). If, in an asymptotically flat spacetime, the semi-Killing vectors are asymptotically Killing, the surface integrals (\ref{CCGR-Komar3Form}) may still be used to define conserved charges---in particular the mass and angular momentum\footnote{Up to an anomalous factor of two \cite{Katz1985} (also see \cite{Lynden-BellBicak2017}, which relates this factor to the trace of the energy-momentum tensor).} for an asymptotically flat spacetime. Another approach, introduced by Harte in \cite{Harte2008}, constructs Komar currents from affine collineations (which form another generalization for Killing vectors), defined as solutions to the equation $\nabla_\alpha \nabla_{(\mu} \xi_{\nu)} = 0 $. 

Here, I describe a generalization of Komar's approach by dropping the divergence-free constraint. One may recognize that Eq. (\ref{CCGR-KillingWaveEquation}) is a curved spacetime wave equation for a vector field; solutions of Eq. (\ref{CCGR-KillingWaveEquation}) form a natural generalization for Killing vectors which in principle can be constructed in a generic spacetime. A further generalization of Eq. (\ref{CCGR-KillingWaveEquation}) is the almost Killing equation (AKE) \cite{Taubes1978,Bonaetal2005,Ruizetal2014}
\begin{equation} \label{CCGR-ApproximateKillingEquation}
\Box \xi^{\nu} + R{^\nu}{_\sigma} \, \xi^{\sigma} + (1 - \mu) \nabla^\nu \left(\nabla \cdot \xi \right) = 0 ,
\end{equation}

\noindent where $\mu$ is a constant parameter. The solutions $\xi^\mu$ of the AKE are termed almost-Killing vectors. The AKE (\ref{CCGR-ApproximateKillingEquation}) is a generalization of Eq. (\ref{CCGR-KillingWaveEquation}), and it is straightforward to verify that if $\xi^\mu$ is a Killing vector, it satisfies the AKE. If $\xi^\mu$ is a solution to the AKE (\ref{CCGR-ApproximateKillingEquation}), the Komar current [Eq. (\ref{CCGR-Komar})] takes the following form:\footnote{One might attempt to exploit the gauge freedom $\xi^\mu \rightarrow \xi^\mu +\nabla^\mu \sigma$ in Eq. (\ref{CCGR-Komar}) for the Komar current to eliminate the divergence of $\xi$, but since the result (\ref{CCGR-KomarAK}) depends on Eq. (\ref{CCGR-ApproximateKillingEquation}), one cannot replace $\xi^\mu$ in the result with its gauge-transformed counterpart in Eq. (\ref{CCGR-KomarAK}); if one attempts to impose a Lorenz gauge $\nabla \cdot \xi = -\Box \sigma$, the most one can do is to replace $\nabla \cdot \xi$ with $- \Box \sigma$ in Eq. (\ref{CCGR-KomarAK}).}
\begin{equation} \label{CCGR-KomarAK}
\begin{aligned}
J^\nu_{AK} &= 2R^{\nu}{_\sigma} \xi^\sigma + (2-\mu) \nabla^\nu \left(\nabla \cdot \xi \right).
\end{aligned}
\end{equation}

\noindent Note that for $\mu=2$, the current vanishes for vacuum spacetimes, and that for $\mu \neq 2$, it measures the degree to which $\nabla \cdot \xi$ fails to be constant. Since Eq. (\ref{CCGR-KomarAK}) is a consequence of Eq. (\ref{CCGR-Komar}), the divergence-free property for the current $J^\mu_{AK}$ (Eq. (\ref{CCGR-KomarDivergence})) is an identity; it follows that if solutions of Eq. (\ref{CCGR-ApproximateKillingEquation}) exist, they must satisfy the following condition:
\begin{equation} \label{CCGR-KomarAKcond}
 \xi^\mu \nabla_\mu R + 2 R^{\mu \nu} \nabla_{(\mu} \xi_{\nu)} + (2-\mu) \Box \left(\nabla \cdot \xi \right)=0.
\end{equation}
 
\noindent I stress that Eq. (\ref{CCGR-KomarAKcond}) should be viewed as a property of solutions for the AKE, not a constraint; it is a consequence of the fact that the Komar current, as defined in  Eq. (\ref{CCGR-Komar}), identically satisfies the divergence-free condition (\ref{CCGR-KomarDivergence}). 

I must mention the globally conserved current of Ruiz et al. \cite{Ruizetal2014}, which is also constructed from almost-Killing vectors. In particular Ruiz \textit{et al.} construct a conserved current from solutions $\xi^\mu$ of the AKE (\ref{CCGR-KomarAK}) by directly generalizing Eq. (\ref{CCGR-KomarKilling}) to obtain the expression $J^\mu_\chi = 2 \, R^\mu{_\nu} \xi^\nu$. Their result is not in general a Komar current, and is not identically divergence-free; one only has a divergence-free current with the choice $\mu=2$ (in which case one has a Komar current) or by imposing the constraint $\Box (\nabla \cdot \xi)=0$. 

Another Komar current may be constructed from the related notion of an approximate Killing vector by Matzner in \cite{Matzner1968} (see also \cite{Beetle2008,*BeetleWilder2014}), which may be of use for constructing Komar currents in compact Riemannian manifolds of positive definite signature. This approach seeks an extremal value for the functional:
\begin{equation} \label{CCGR-ApproximateKillingEquationMatzner}
\lambda[\xi]:=\frac{\int \nabla^{(\mu} \xi^{\nu)} \nabla_{(\mu} \xi_{\nu)} dV}{\int \xi^\mu \xi_\mu dV} ,
\end{equation}

\noindent where $dV=\sqrt{|g|} d^n x$ is the volume element for the manifold. Extremizing the functional $\lambda[\xi]$ yields the following equation:
\begin{equation} \label{CCGR-ApproximateKillingEquationMatzner}
\begin{aligned}
	\Delta \xi_\mu := & \nabla^\nu \left(\nabla_\mu \xi_\nu + \nabla_\nu \xi_\mu \right) = \lambda \xi_\mu 
,
\end{aligned}
\end{equation}

\noindent which one recognizes to be an eigenproblem for the operator $\Delta$. The resulting Komar current is:
\begin{equation} \label{CCGR-ApproximateKillingEquationCurrent}
\begin{aligned}
J^\mu_{M} 
	&= 2 R^{\mu}{_\nu} \xi^\nu+ 2\nabla^\mu (\nabla \cdot \xi) - \lambda \xi^\mu.
\end{aligned}
\end{equation}

\noindent In compact Riemannian manifolds (with positive-definite signature),  the spectrum of eigenvalues $\lambda$ is discrete and nonnegative, and the eigenvalue $\lambda = 0$ is contained in the spectrum if and only if the corresponding eigenvector is a Killing vector \cite{Matzner1968}. Approximate Killing vectors in such manifolds may therefore be defined as solutions to Eq. (\ref{CCGR-ApproximateKillingEquationMatzner}) which have the minimum value for the eigenvalue $\lambda$. 

Unfortunately, these properties no longer hold for eigensolutions on noncompact spacetimes with Lorentzian signature, and the eigenvalue $\lambda=0$ is no longer unique. Nevertheless, the approximate Killing vector approach still provides some useful insights; Matzner demonstrates that the functional $\lambda[\xi]$ provides a measure of the deviation from symmetry for linearized gravitational waves, and in the context of the averaging procedure of Isaacson \cite{Isaacson1968a,*Isaacson1968b} for high frequency gravitational radiation, the functional $\lambda[\xi]$ depends on the averaged effective energy-momentum tensor for gravitational radiation \cite{Matzner1968}.\footnote{An interesting line of investigation, which I leave for future work, concerns the precise relationship between $\lambda[\xi]$ and the conserved charges for the Komar current calculated from the eigenvectors $\xi^\mu$.}  

I also note that for Ricci-flat manifolds, the values of the current $J^\mu_{M}$ (\ref{CCGR-ApproximateKillingEquationCurrent}) provides (through $\lambda$) a nonlocal measure of the degree to which $\xi^\mu$ fails to be a Killing vector. On the other hand, the current $J^\mu_{AK}$ (\ref{CCGR-KomarAK}) constructed only measures the degree to which the $\nabla \cdot \xi$ fails to be constant. The current $J^\mu_{M}$ (\ref{CCGR-ApproximateKillingEquationCurrent}) constructed from eigensolutions of (\ref{CCGR-ApproximateKillingEquationMatzner}) therefore provides a more complete\footnote{Though Matzner points out that on manifolds of Lorentzian signature, the tensor $\nabla_{(\mu} \xi_{\nu)}$ can be null ($\nabla_{(\mu} \xi_{\nu)} \nabla^{(\mu} \xi^{\nu)}=0$), even if $\xi^\mu$ is not null \cite{Matzner1968}, so $\lambda[\xi]$ does not provide a fully complete measure of the degree to which $\xi^\mu$ fails to be Killing.} measure of the degree to which $\xi^\mu$  fails to satisfy Killing's equation.



\section{An Example: The Vaidya Spacetime}
The conserved currents presented in the preceding section yield conserved quantities that may be thought of as generalizations of energy and momentum. Of course, the generalized Killing vectors on which these currents are based are by no means unique (I shall construct more later on in this article), so one cannot regard the corresponding conserved quantities as definitions for the true energy and momentum of a gravitating system. Nonetheless, the conserved quantities constructed from the conserved currents presented in this article may be useful as phenomenological (and local) definitions for conserved energy- and momentumlike quantities.

Here, I illustrate this for the outgoing Vaidya spacetime, which describes the spacetime geometry in the exterior of a radiating object; the line element for the Vaidya spacetime is (I set $G=1$ throughout this article):
\begin{equation}\label{CCGR-VaidyaLineElem}
ds^2=-\left(1-\frac{2 M(u)}{r}\right) du^2 - 2 du dr + r^2 \left(d\theta^2+\sin^2 \theta \, d\phi^2\right) ,
\end{equation}

\noindent and I consider the case where the mass function $M(u)$ has the form:
\begin{equation}\label{CCGR-MassFunction}
M(u)=M_{av}-\tfrac{1}{2}\delta M \, \text{erf}(\alpha \, u) ,
\end{equation}

\noindent where $\text{erf}(x):=2/\sqrt{\pi} \int^x_0 e^{-q^2} dq$ is the error function and $\alpha$ is a constant parameter. This mass function describes the spacetime around a spherical object, initially of mass $M_{av}+\delta M/2$, that emits a uniform pulse of radiation and loses a mass $\delta M$ in the process. For $\mu=2$, the AKE (\ref{CCGR-ApproximateKillingEquation}) admits a solution of the following form:
\begin{equation} \label{CCGR-AKEVaidyaPulse}
\xi = \left(1-\frac{\delta M \text{erf}(\alpha u)}{2M},-\frac{\delta M \alpha r e^{-\alpha^2 u^2}}{\sqrt{\pi}M},0,0\right).
\end{equation}

\noindent The Komar current (\ref{CCGR-KomarAK}) for $\xi$ is then:
\begin{equation} \label{CCGR-AKEVaidyaPulseKomar}
J_{AK}= \left(0,-\frac{2 \delta M \alpha e^{-\alpha^2 u^2} \left(2M-\delta M\text{erf}(\alpha u)\right)}{\sqrt{\pi}M r^2},0,0\right).
\end{equation}

\noindent On a spacelike constraint surface $\Sigma_s$ defined by the constraint function $s(u,r)=u+r$, I can perform the following integral (I have made use of $u=s-r$, with $s$ constant over $\Sigma_s$) over the domain $R_1<r<R_2$ to obtain the following energylike quantity:\footnote{The factor of $1/8\pi$ in front of the integral in Eq. (\ref{CCGRA-ConservedIntegralVaidya}) has been chosen so that one recovers $Q=M$ for the case of a static Schwarzschild spacetime.}
\begin{equation} \label{CCGRA-ConservedIntegralVaidya}
\begin{aligned}
Q =&-\frac{1}{8\pi}\int_{\Sigma_{s}} n_\mu \, J^\mu_{AK} \, \sqrt{|h|} d^3y \\
    = & \frac{\delta M^2}{4M} \biggl(\text{erf}(\alpha  (s-R_2))^2 - \text{erf}(\alpha  (s-R_1))^2\biggr)\\
& + \delta M\biggl(\text{erf}(\alpha  (s-R_1)) - \text{erf}(\alpha (s-R_2))\biggr) .
\end{aligned}
\end{equation}

\noindent Since $\Sigma_s$ is spacelike, $s$ is effectively a time coordinate, and the pulse is centered at the radius $r=s$. One can see that when the pulse is well contained in the domain $R_1<r<R_2$ (in particular, for $s \gg R_1$ and $s \ll R_2$), $Q$ is nearly independent of $s$, and has a value $Q \sim 2 \delta M$ (note that I am evaluating this on a finite domain, which is why there is no term containing $M$). Also note that the first term nearly vanishes when the pulse is far from $R_1$ and $R_2$. When the pulse is far outside the domain $R_1<r<R_2$, $s \gg R_2$, so the value of $Q$ decreases to zero. The quantity $Q$ therefore behaves in a manner expected for the energy contained in a uniform pulse of radiation emitted by a spherical object, except that the change in the charge $\delta Q \sim 2 \delta M$ is twice the expected value. The factor of two here comes from the fact that the energy-momentum tensor for the Vaidya spacetime is trace-free; it was noted in \cite{Lynden-BellBicak2017} that Komar integrals for trace-free energy-momentum tensors yield values twice that of energy-momentum tensors with nonvanishing trace. The results I obtain here further establishes those of \cite{Lynden-BellBicak2017}.


\section{Komar current from conformal Killing vectors and their generalization}
I now consider the construction of the Komar current from a conformal Killing vector and its generalization to generic spacetimes. A conformal Killing vector is defined as a vector $\xi$ which satisfies the following \cite{Wald}:
\begin{equation} \label{CCGR-KillingsEquationConformal}
\pounds_\xi g_{\mu \nu}=2\nabla_{(\mu} \xi_{\nu)} = \psi \,  g_{\mu \nu},
\end{equation}

\noindent where $\psi$ is a scalar field. Note that $\nabla_\mu \xi^\mu = 2 \, \psi$. Taking the divergence of (\ref{CCGR-KillingsEquationConformal}) yields:
\begin{equation} \label{CCGR-KillingsEquationConformalWave}
\begin{aligned}
	&\Box \xi_{\mu} +R{_{\mu\nu}}\xi^{\nu} = - \nabla_\mu \psi .
\end{aligned}
\end{equation}

\noindent I now turn to the Komar current $J^\mu_K$ [Eq. (\ref{CCGR-Komar})], which now takes the form:
\begin{equation} \label{CCGR-ConformalKomar}
\begin{aligned}
J^\mu_C 
	&= 2 \, R^\mu{_\nu} \xi^\nu + 3 \nabla^\mu \psi .
\end{aligned}
\end{equation}

\noindent Note that in a vacuum, the value of $J^\mu_C$ measures the degree to which $\xi^\mu$ fails to be homothetic ($\psi=\text{const.}$). The Komar identity $\nabla_\mu J^\mu_K = 0$ demands:
\begin{equation} \label{CCGR-KomarIDConformal}
\begin{aligned}
\nabla_\mu J^\mu_C  
	&= \xi^\nu \nabla_\nu R  + R \, \psi + 3 \Box \psi = 0,
\end{aligned}
\end{equation}

\noindent so that the conformal factor $\psi$ satisfies the following wave equation:
\begin{equation} \label{CCGR-ConformalWaveEq}
\Box \psi+ \frac{1}{3} \, R \, \psi = - \frac{1}{3} \xi^\nu \nabla_\nu R .
\end{equation}

\noindent Since Eq. (\ref{CCGR-ConformalWaveEq}) is derived from an identity $\nabla_\mu J^\mu_K=0$, this result demonstrates that if a conformal Killing vector exists, its associated conformal factor $\nabla_\mu \xi^\mu = 2 \, \varphi$ must satisfy the wave equation (\ref{CCGR-ConformalWaveEq}).

It is worth pointing out that in simple cosmological spacetimes, the conformal Komar current in Eq. (\ref{CCGR-ConformalKomar}) vanishes. It is well known that the Friedmann--Lema{\^i}tre--Robertson--Walker (FLRW) metric admits a conformal timelike Killing vector, and it is natural to construct a Komar current from this vector. The FLRW spacetime may be described by the line element:\footnote{I set $c=1$ and employ the MTW \cite{MTW} signature $\left(-,+,+,+\right)$.}
\begin{equation} \label{CCGR-FLRWline}
%
	ds^2 =a^2(\tau) \left( - d\tau^2 + \frac{dr^2}{1- k \, r^2} + r^2 \left(d\theta^2 + \sin^2 \theta \, d\phi^2 \right) \right),
\end{equation}

\noindent where the conformal time coordinate $\tau$ is related to the usual comoving time coordinate $t$ by $dt = a(\tau) d\tau$. It is straightforward to see that the line element (\ref{CCGR-FLRWline}) admits a timelike conformal Killing vector; the only part of the line element dependent on the coordinate $\tau$ is the scale factor $a(\tau)$, which only appears once as a conformal factor. For the FLRW metric, the Lie derivative is proportional to the metric:
\begin{equation} \label{CCGR-FLRWLieConformal}
\pounds_{\frac{\partial}{\partial \tau}} g_{\mu \nu} = 2 \frac{\dot{a}(\tau)}{a(\tau)} \, g_{\mu \nu} .
\end{equation}

\noindent The coordinate basis vector $\partial /\partial \tau$ is therefore a conformal Killing vector.

Since the FLRW admits a timelike conformal Killing vector $\partial /\partial \tau$, it is natural to construct a Komar current from $\partial /\partial \tau$. Setting $x^0 = \tau$, the components of $\partial /\partial \tau$ are $\delta^\mu_0$, with $\delta^\mu_\nu$ being the Kronecker delta. The Komar current that results from this vanishes; to see this, note that the metric component $g_{\mu 0}$ form the covariant components of the vector $\delta^\mu_0$. In terms of $g_{\mu 0}$, the conformal Komar current is:
\begin{equation} \label{CCGR-FLRWConformalKomarCurrent}
J^\mu_C = g^{\mu \alpha}\nabla^\beta \left( \partial_\alpha g_{\beta 0} - \partial_\beta g_{\alpha 0} \right).
\end{equation}

\noindent The quantity in the brackets vanishes because the metric is diagonal and $g_{00}$ depends only on $\tau$. The Komar current vanishes, and it follows that the resulting Komar integrals also vanish.\footnote{This applies to the surface integrals in Eq. (\ref{CCGR-Komar3Form}); note that the integrand of the surface integrals depend on the bracketed quantity in Eq. (\ref{CCGR-FLRWConformalKomarCurrent}).} It is not surprising to find that the Komar current and its associated charges vanish; the FLRW spacetime possesses a high degree of symmetry, and for a closed universe, the vanishing of charges immediately follows from Eq. (4).

The conformal Komar current motivates the construction of a more general almost-conformal Komar current, valid in spacetimes which do not admit a conformal Killing vector. In principle, one can construct Eq. (\ref{CCGR-KillingsEquationConformalWave}) in a generic spacetime:
\begin{equation} \label{CCGR-ConformalKillingWaveEquation}
\Box \zeta_{\mu} + R_{\mu \nu} \zeta^{\nu} = - \nabla_\mu \omega .
\end{equation}

\noindent The Komar current for $\zeta^\mu$ satisfying Eq. (\ref{CCGR-ConformalKillingWaveEquation}) is
\begin{equation} \label{CCGR-GenConformalKomarCurrent}
\begin{aligned}
J^\mu_{AC}
	&= 2 \,R^\mu{_\nu} \zeta^\nu + \nabla^\mu (\nabla \cdot \zeta) + \nabla^\mu \omega .
\end{aligned}
\end{equation}

\noindent The divergence of the above yields:
\begin{equation} \label{CCGR-KomarIDGenConformal}
\begin{aligned}
	\zeta^\nu \nabla_\nu R  + 2 \, R^{\mu \nu} \nabla_{(\mu} \zeta_{\nu)} + \Box (\nabla \cdot \zeta) + \Box \omega = 0. 
\end{aligned}
\end{equation}

\noindent Now since the divergence of the Komar current is identically zero, Eq. (\ref{CCGR-KomarIDGenConformal}) does not constrain the scalar field $\omega$; Eq. (\ref{CCGR-KomarIDGenConformal}) is a property of any solution $\zeta^\mu$ to Eq. (\ref{CCGR-ConformalKillingWaveEquation}). Since I intend $\zeta^\mu$ to be a generalization of the conformal Killing vector, I require that $\omega$ satisfies Eq. (\ref{CCGR-ConformalWaveEq}):
\begin{equation} \label{CCGR-GenConformalWaveEq}
\Box \omega + \frac{1}{3} \, R \, \omega = - \frac{1}{3} \zeta^\nu \nabla_\nu R .
\end{equation}

\noindent One may then solve Eqs. (\ref{CCGR-ConformalKillingWaveEquation}) and (\ref{CCGR-GenConformalWaveEq}) for $\zeta^\mu$ and $\omega$; the resulting vector field $\zeta^\mu$ may then be used to construct a Komar current in a generic spacetime. If they exist, Killing vectors and conformal Killing vectors both lie in the solution space of Eqs. (\ref{CCGR-ConformalKillingWaveEquation}) and (\ref{CCGR-GenConformalWaveEq}). If $\xi^\mu$ is a Killing vector, then Eq. (\ref{CCGR-ConformalKillingWaveEquation}) requires that $\omega$ be a constant, and Eq. (\ref{CCGR-GenConformalWaveEq}) implies that $\omega$
must vanish. If $\zeta^\mu$ is a conformal Killing vector, then Eq. (\ref{CCGR-ConformalKillingWaveEquation}) requires that up to a constant, $2\omega = \nabla \cdot \zeta$, and the wave equation for $\omega$ (Eq. (\ref{CCGR-GenConformalWaveEq})), combined with the property (\ref{CCGR-KomarIDGenConformal}), requires that $2\omega = \nabla \cdot \zeta$ holds exactly. The solutions of Eqs. (\ref{CCGR-ConformalKillingWaveEquation}) and (\ref{CCGR-GenConformalWaveEq}) therefore provide a suitable generalization for both Killing vectors and conformal Killing vectors in generic spacetimes. For situations in which one has an approximate conformal symmetry, it may be appropriate to formulate conservation laws using the Komar current (\ref{CCGR-GenConformalKomarCurrent}) constructed from solutions of Eqs. (\ref{CCGR-ConformalKillingWaveEquation}) and (\ref{CCGR-GenConformalWaveEq}).


\section{Globally conserved non-Komar currents from scalar test fields}
As mentioned earlier, the Komar current has theoretical appeal because it can be derived from fundamental principles in the context of N{o}ether's theorem \cite{Katz1985,Baketal1993,Obukhovetal2006a,*Obukhovetal2006b}. However, Komar currents may be of limited use in certain circumstances. As discussed earlier, any globally conserved charge constructed from the Komar current will vanish in spatially closed universes, by virtue of Eq. (\ref{CCGR-Komar3Form}). Even for Schwarzschild and Kerr spacetimes, which admit a nonvanishing Komar charge for timelike Killing vectors, the Komar current $J^\mu_\chi$ [Eq. (\ref{CCGR-KomarKilling})] for Killing vectors is of little use locally since it vanishes in vacuum spacetimes. While the currents constructed from generalized Killing vectors in Eqs. (\ref{CCGR-KomarAK}) and (\ref{CCGR-GenConformalKomarCurrent}) do not in general vanish in vacuum spacetimes, and their nonvanishing values measure the degree to which the divergence of the generalized Killing vector fails to be constant---it would be preferable instead to have a current which more completely measures the degree to which the generalized Killing vector fails to satisfy Killing's equation. Though the current Eq. (\ref{CCGR-ApproximateKillingEquationCurrent}) constructed from Matzner's approximate Killing vector can be interpreted as such, it depends on $\lambda[\xi]$ which, being constructed from integrals, is difficult to evaluate on the whole of a noncompact spacetime.

Fortunately, one can construct globally conserved currents that do not correspond to a Komar current---the conserved current of Ruiz \textit{et al.} is one example \cite{Ruizetal2014} (note, however, that it also vanishes in a vacuum spacetime). There is a more general class of conserved currents (containing the Komar current) defined as those formed from divergences of superpotentials; the discussion of this approach is beyond the scope of this article, and I refer the reader to \cite{PetrovKatz2002,*KatzLivshits2008,*JuliaSilva1998,KBL1997,Szabados2009} and references contained therein. In this section, I examine a simple construction that can turn a nonconserved current into a globally conserved current in a generic spacetime. As I shall demonstrate, the simplicity of this construction facilitates both the computation and interpretation of the resulting currents. 

To motivate this construction, recall that the identity $\nabla_\mu J^\mu_K=0$ for the Komar current $J^\mu_K$ (\ref{CCGR-KomarDivergence}) admits a shift freedom; one can add any divergence-free vector to the Komar current $J^\mu_K$ to obtain another conserved current. In particular, I note that the current $J^\mu_K + \nabla^\mu \varphi$ is also divergence-free when the scalar field $\varphi$ satisfies the homogeneous wave equation $\Box \varphi = 0$. In FLRW spacetime, one solution to the homogeneous wave equation is $\varphi = - \int \left(\rho_0 /a^2(\tau) \right) d\tau$ (where $\rho_0$ is a constant); this produces a shift in the Komar charge by an amount $\rho_0 \, V$ where $V$ is the spatial volume when $a(\tau)=1$. This property is reminiscent of the shift freedom in the electrostatic potential or the Newtonian gravitational potential.\footnote{In the gravitational case, note that for a uniform gravitational field (like that near the surface of the earth) there is no absolute definition for the potential energy; the shift freedom corresponds to a freedom of choice in the reference height.}

Given a current $J^\mu$ that is not divergence free ($\nabla_\mu J^\mu \neq 0$), I may construct a conserved current with a similar procedure---I add to $J^\mu$ a gradient $\nabla^\mu$ which cancels out the divergence:
\begin{equation} \label{CCGR-PhenomenologicalCurrentGeneral}
K^\mu = J^\mu - \nabla^\mu \varphi ,
\end{equation}

\noindent where the scalar field $\varphi$ now satisfies an inhomogeneous wave equation:
\begin{equation} \label{CCGR-PhenomenologicalCurrentGeneralWave}
\Box \varphi = \nabla \cdot J.
\end{equation}

\noindent In this framework, the scalar field $\varphi$ absorbs the divergences for the current $J^\mu$, and characterizes the degree to which the current $J^\mu$ and its charges fail to satisfy conservation laws. Of course, Eq. (\ref{CCGR-PhenomenologicalCurrentGeneralWave}) admits many more solutions than necessary for this purpose,\footnote{Here, I assume that for a given set of initial conditions on a Cauchy surface (and a well-behaved source), unique solutions to (\ref{CCGR-PhenomenologicalCurrentGeneralWave}) exist for a sufficiently short (but finite) time.} since one can add solutions of the homogeneous wave equation $\Box \varphi = 0$ to a particular solution of Eq. (\ref{CCGR-PhenomenologicalCurrentGeneralWave}). These additional solutions correspond to shifts in the conserved charges (recall the discussion in the preceding paragraph for the FLRW example). This can be made explicit for the current $J^\mu = R^{\mu \nu} n_\nu$, where $n^\mu$ is the unit normal vector to constant $t$ surfaces:
\begin{equation} \label{CCGR-PhenomenologicalCurrent1}
K^\mu_{R} = R^{\mu \nu} \, n_\mu - \nabla^\mu \varphi.
\end{equation}

\noindent The resulting charge $Q$ may be interpreted as an energy for an FLRW spacetime. For a closed FLRW spacetime with a scale factor\footnote{Here, I use the usual comoving coordinate $t$, in which the line element takes the form $ds^2=-dt^2+a^2(t) d\mathcal{S}^2$, where $d\mathcal{S}^2$ is the line element for constant $t$ surfaces when $a(t)=1$.} $a(t)=a_0 t^q$, I obtain the following solution of Eq. (\ref{CCGR-PhenomenologicalCurrentGeneralWave}) for $\varphi=\varphi(t)$:
\begin{equation} \label{CCGR-ScalarSolutionFLRW}
\varphi(t) = \frac{3 q (q-1)}{t} + C_1 \frac{t^{1-3q}}{1-3q} +C_2 ,
\end{equation}

\noindent where $C_1$ and $C_2$ are constants of integration. The conserved charge (evaluated on a constant $t$ surface) is then:
\begin{equation} \label{CCGR-ScalarSolutionFLRW}
Q = a_0^3 C_1 V ,
\end{equation}

\noindent where $V$ is the volume of the constant $t$ hypersurface when $a(t)=1$. Note that $Q$ depends on the constant of integration $C_1$; it is straightforward to verify that the term containing $C_1$ in Eq. (\ref{CCGR-ScalarSolutionFLRW}) is in fact a solution of the homogeneous wave equation $\Box \varphi$, which generates a shift in the energy. If I set $C_1=0$, I recover the result that the conserved charge (which corresponds to an energy) vanishes; the $C_1=0$ solution therefore yields the correct value (zero) for the total energy in a closed FLRW universe. This example further establishes that solutions to the homogeneous equation correspond to shifts in the conserved charges.

I now discuss a current closely related to $K^\mu_{R}$ (\ref{CCGR-PhenomenologicalCurrent1}), which endows the scalar field $\varphi$ with an interesting property in a spatially closed (not necessarily FLRW) universe. The current takes the following form:
\begin{equation} \label{CCGR-PhenomenologicalCurrent2}
K^\mu_{T} = T^{\mu \nu} \, X_\nu - \nabla^\mu \varphi.
\end{equation}

\noindent where $X^\mu$ is a timelike, future-pointing unit vector, and $T^{\mu \nu}$ is an energy-momentum tensor that satisfies the dominant energy condition \cite{HawkingEllis,Wald}, which for the discussion here takes the form $T_{\mu \nu} X^\mu Z^\nu \geq 0$ for any two future-directed timelike unit vectors $X^\mu$ and $Z^\mu$. I note that one can always choose initial conditions for $\varphi$ such that (where $n^\mu$ is a timelike unit normal vector to a constant $t$ hypersurface):
\begin{equation} \label{CCGR-Scalar2Initial}
n^\mu \nabla_\mu \varphi - T_{\mu \nu} n^\mu X^\nu = 0 ,
\end{equation}

\noindent so that the resulting conserved charge takes the value $Q = 0$. Equation (\ref{CCGR-PhenomenologicalCurrentGeneralWave}) ensures that $K^\mu_{T}$ is divergence-free so that the value of the charge $Q$ is foliation independent. If $Q=0$, it follows that on any hypersurface $\Sigma$:
\begin{equation} \label{CCGR-ChargeCond}
\int_{\Sigma} n^\mu \nabla_\mu \varphi \sqrt{|h|} d^3 x \geq 0 ,
\end{equation}

\noindent where $h$ is the determinant of the induced metric on $\Sigma$. Since the above inequality holds for any spacelike hypersurface, I find that the dominant energy condition implies that for any future-directed hypersurface-orthogonal vector $Z^\mu$, the integral of the derivative $Z^\mu \nabla_\mu \varphi$ on the integral hypersurfaces of $Z^\mu$ is positive. This condition is the statement that on average, $\varphi$ is constant or increasing\footnote{I should point out that since the volume of $\Sigma$ differs between hypersurfaces, the inequality in Eq. (\ref{CCGR-ChargeCond}) does not imply that the integral $\int_{\Sigma} \varphi \sqrt{|h|} d^3 x$ increases for uniform infinitesimal displacements in the direction of $n^\mu$. This is only true on maximal hypersurfaces (hypersurfaces which have vanishing mean curvature).} along future-directed integral curves of $Z^\mu$. I emphasize that this property is only true on average, since the inequality (\ref{CCGR-ChargeCond}) does not in general imply $Z^\mu \nabla_\mu \varphi<0$ at every point in $\Sigma$. Nevertheless, this property suggests that $\varphi$ is globally increasing along the integral curves of a unit hypersurface-orthogonal timelike vector field $Z^\mu$. The current (\ref{CCGR-PhenomenologicalCurrent2}) may therefore provide a framework for studying questions concerning the arrow of time and its relationship to energy conditions in general relativity.

Finally, I discuss a construction that generalizes the Komar current $J^\mu_{AK}$ for almost Killing vectors:\footnote{One may instead substitute the more general current $J^\mu_{AC}$ in Eq. (\ref{CCGR-GenConformalKomarCurrent}), but similar arguments apply---for simplicity, I restrict my attention to $J^\mu_{AK}$.} 
\begin{equation} \label{CCGR-PhenomenologicalCurrentAK}
K^\mu_{AK} = J^\mu_{AK} + \xi_\nu \left( 2 \nabla^{(\mu} \xi^{\nu)} + \kappa g^{\mu \nu} \left( \nabla \cdot \xi \right) \right) - \nabla^\mu \varphi,
\end{equation}

\noindent where $\kappa$ is a parameter and $\varphi$ satisfies:
\begin{equation} \label{CCGR-PhenomenologicalCurrentwave}
\begin{aligned}
\Box \varphi = \nabla_{\mu}  \left(\xi_{\nu} \left( 2 \nabla^{(\mu} \xi^{\nu)} + \kappa g^{\mu \nu} \left( \nabla \cdot \xi \right) \right) \right).
\end{aligned}
\end{equation}

\noindent The current $J^\mu_{AK}$ does not contribute to the right-hand side of (\ref{CCGR-PhenomenologicalCurrentwave}) because it is a Komar current and is identically conserved. Note that when $\xi^\mu$ is a Killing vector, $K^\mu_{AK}$ becomes $J^\mu_\chi$ [Eq.(\ref{CCGR-KomarKilling})] after eliminating the superfluous solutions of the wave equation for $\varphi$ (which may be done with an appropriate choice of initial and boundary conditions). I also point out that the scalar field $\varphi$ may not always be necessary to ensure conservation; if $\kappa =-2 \mu$ and if $\nabla_{(\mu} \xi_{\nu)}$ is null and trace-free,\footnote{Again, as pointed out earlier in footnote 7, the tensor $\nabla_{(\mu} \xi_{\nu)}$ can be null, even for non-Killing vectors $\xi^mu$ that are not themselves null \cite{Matzner1968}} the rhs of (\ref{CCGR-PhenomenologicalCurrentwave}) vanishes and $\varphi$ decouples from $\xi$.

In general, the current $K^\mu_{AK}$ is nonvanishing in vacuum spacetimes which do not admit Killing vectors. While this is true to some degree for the currents in Eqs. (\ref{CCGR-KomarAK}), (\ref{CCGR-ApproximateKillingEquationCurrent}) and (\ref{CCGR-GenConformalKomarCurrent}), the expression has a clear interpretation in the $\mu=2$ case (in which $J^\mu_{AK}$ vanishes for a vacuum spacetime); the terms explicitly dependent on $\xi^\mu$ provide a local measure of the degree to which $\xi^\mu$ fails to satisfy Killing's equation, and the inhomogeneous solutions $\varphi$ of Eq. (\ref{CCGR-PhenomenologicalCurrentwave}) provide a nonlocal measure of the degree to which $\xi^\mu$ fails to be Killing. The current $K^\mu_{AK}$ therefore provides an energy- and momentum-like measure of the deviation from symmetry in a generic spacetime. Note also that the computation of $K^\mu_{AK}$ is much simpler than that of Eq. (\ref{CCGR-ApproximateKillingEquationCurrent}), and can be formulated as an initial value problem for the AKE (\ref{CCGR-ApproximateKillingEquation}) and the wave equation for $\varphi$ (\ref{CCGR-PhenomenologicalCurrentwave}).



\section{Outlook}
In this article, I have constructed some Komar currents [Eqs. (\ref{CCGR-KomarAK}), (\ref{CCGR-ApproximateKillingEquationCurrent}), and (\ref{CCGR-GenConformalKomarCurrent})] from various generalizations of Killing vectors, defined as the respective solutions to Eqs. (\ref{CCGR-ApproximateKillingEquation}), (\ref{CCGR-ApproximateKillingEquationMatzner}), and the system given by Eqs. (\ref{CCGR-ConformalKillingWaveEquation}) and (\ref{CCGR-GenConformalWaveEq}), which can in principle be constructed in generic spacetimes. In spacetimes that admit Killing vectors, I have shown that Killing vectors lie in the solution space to these equations, and Killing vector solutions may be recovered with an appropriate choice of initial and boundary conditions\footnote{In the case of Eq. (\ref{CCGR-ApproximateKillingEquationMatzner}), one must also select the $\lambda=0$ eigensolutions.}---and I have also argued that for Killing vector solutions, the corresponding currents reduce to the familiar Komar currents for Killing vectors. Though the analysis here of these currents and their properties is admittedly a cursory one, I have included a simple example for the Vaidya spacetime which demonstrates how Komar currents from approximate Killing vectors can be used to define conserved quantities that behave in a manner expected for the energy (up to a factor of 2) contained in the outgoing radiation. A more detailed investigation, which I leave for future work, will involve the further study of the systems (in particular their solutions in various spacetime geometries) described in Eqs. (\ref{CCGR-ApproximateKillingEquation}), (\ref{CCGR-ApproximateKillingEquationMatzner}) and Eqs. (\ref{CCGR-ConformalKillingWaveEquation}), (\ref{CCGR-GenConformalWaveEq}).

I have also presented a new class of globally conserved currents which do not correspond to Komar currents, and can be constructed from an existing current (which does not need to be conserved) and a scalar field. While these currents do not have the same fundamental status as Komar currents (which can be derived in the framework of Noether's theorem), they are simple to construct, and I have shown that they can have interesting features which may be of conceptual and calculational utility. I discussed a couple of examples; the current (\ref{CCGR-PhenomenologicalCurrent2}) is constructed from a unit timelike vector and the energy-momentum tensor, and I have shown that under the dominant energy condition, the scalar field $\varphi$ is on average constant or increasing in the direction of future-pointing hypersurface orthogonal timelike unit vectors---this feature might make this construction a useful framework for studying questions concerning the arrow of time and its possible relationship to energy conditions in general relativity.  The current (\ref{CCGR-PhenomenologicalCurrentAK}) is constructed from almost-Killing vectors and (with appropriate initial conditions) yields a globally conserved energy- and momentum-like quantity which measures the degree to which a given spacetime deviates from symmetry. The examples discussed in this article are by no means unique; there are many more currents that one can construct from the prescription in Eqs. (\ref{CCGR-PhenomenologicalCurrentGeneral}) and (\ref{CCGR-PhenomenologicalCurrentGeneralWave}). I leave for future investigation the exploration and identification of other useful currents that may be constructed with this prescription.



\begin{acknowledgments}
This work was partially supported by the National Science Foundation under Grant No. PHY-1620610. I thank Richard A. Matzner for his comments and feedback on this work.
\end{acknowledgments}


\appendix*

\section{Killing Vectors and Conservation Laws}\label{Appendix-KV}
In this Appendix, I briefly review the relationship between Killing vectors and global conservation laws, the discussion of which may be found in a standard reference on general relativity (see for instance \cite{Poisson,MTW,Weinberg,Wald,Carroll}) or in the seminal work of Komar \cite{Komar1962}. A Killing vector $\chi^\mu$ is defined by Killing's equation:
\begin{equation} \label{CCGRA-KillingsEquation}
\frac{1}{2}\pounds_\chi g_{\mu \nu}=\nabla_{(\mu} \chi_{\nu)}=0 ,
\end{equation}

\noindent where $\pounds_\chi$ is the Lie derivative, and I use the symmetrization convention $A_{(\mu \nu)}=\frac{1}{2} \left(A_{\mu \nu}+A_{\nu \mu}\right)$. From Eq. (\ref{CCGRA-KillingsEquation}), it is straightforward to show that a Killing vector is divergence-free $\nabla_\mu \chi^\mu = 0$, and that $\chi^\mu$ satisfies the following wave equation:
\begin{equation} \label{CCGRA-KillingWaveEquation}
\Box \chi^{\mu} + R{^\mu}{_\nu} \, \chi^{\nu} = 0 .
\end{equation}

\noindent where $R_{\mu \nu}$ is the Ricci tensor. Killing vectors describe the isometries of spacetime; if a metric admits a Killing vector, there exists a coordinate system in which the metric is independent of a coordinate and the Killing vector becomes a coordinate basis vector for that coordinate.\footnote{To see this, recall that the Lie derivative with respect to a coordinate basis vector is just the partial derivative. For a coordinate $x$, this means that $\pounds_{\partial/\partial x}=\partial/\partial x$. Killing's equation for $\partial / \partial x$ then becomes $\partial g_{\mu \nu}/\partial x = 0$, which is precisely the condition that $g_{\mu \nu}$ is independent of the coordinate $x$.}

Given a Killing vector, one may construct a conserved current from the energy-momentum tensor for matter $T^{\mu \nu}$:
\begin{equation} \label{CCGRA-ConservedCurrentT}
J^\mu_T := T^{\mu \nu} \chi_\nu .
\end{equation}

\noindent The divergence of $J^\mu_T$ vanishes:
\begin{equation} \label{CCGRA-ConservedCurrentDivT}
\nabla_\mu J^\mu_T = \nabla_{(\mu} \chi_{\nu)} T^{\mu \nu} + \chi_\nu \, \nabla_\mu T^{\mu \nu} = 0 .
\end{equation}

\noindent The second equality follows from Killing's equation (\ref{CCGRA-KillingsEquation}) and the local conservation law $\nabla_\mu T^{\mu \nu} = 0$ for the energy-momentum tensor. 

One can construct another conserved current from the Ricci tensor $R_{\mu \nu}$:
\begin{equation} \label{CCGRA-ConservedCurrentR}
J^\mu_R := R^{\mu \nu} \chi_\nu .
\end{equation}

\noindent The divergence of $J^\mu_R$ takes the following form:
\begin{equation} \label{CCGRA-ConservedCurrentDivR1}
\nabla_\mu J^\mu_R = \nabla_{(\mu} \chi_{\nu)} R^{\mu \nu} + \chi_\nu \, \nabla_\mu R^{\mu \nu}.
\end{equation}

\noindent The divergence of $R^{\mu \nu}$ does not vanish, but it does satisfy the contracted Bianchi identity:
\begin{equation} \label{CCGRA-ContractedBianchiIdentity}
\nabla_\mu R^{\mu \nu} = \frac{1}{2}\, \nabla^\nu R 
\end{equation}

\noindent The last term in (\ref{CCGRA-ConservedCurrentDivR1}) vanishes by way of $\chi^\nu \nabla_\nu R = 0$; this property follows from the observation that if the Killing vector  $\chi$ is a coordinate basis vector so that $\chi=\partial/\partial z$, then the metric and the Ricci scalar $R$ become independent of the coordinate $z$. It then follows that the divergence of $J^\mu_R$ vanishes
\begin{equation} \label{CCGRA-ConservedCurrentDivR2}
\nabla_\mu J^\mu_R = 0.
\end{equation}

It is straightforward to construct globally conserved quantities from $J^\mu$, provided $\nabla_\mu J^\mu=0$. To do this, integrate $\nabla_\mu J^\mu=0$ over some region of spacetime $U$ (with boundary $\partial U$) and apply the divergence theorem to obtain the result:
\begin{equation} \label{CCGRA-ConservedCurrentDivInt}
\int_{U} \nabla_\mu J^\mu \sqrt{|g|} d^4x = \int_{\partial U} n_\mu \, J^\mu \, \varepsilon \, \sqrt{|h|} d^3y= 0 .
\end{equation}

\noindent where $n^\mu$ is the outward pointing unit normal vector to the boundary $\partial U$, with $\varepsilon=n^\mu\,n_\mu=\pm 1$, $y$ are coordinates on $\partial U$, and $h=\det(h_{ij})$, with $h_{ij}$ being the induced metric on $\partial U$ (here, I assume the $\partial U$ is non-null). If the spacetime manifold has topology $\mathbb{R} \times \Sigma$, where $\Sigma$ is compact and without boundary,\footnote{I make this assumption to simplify the argument, but one can make similar arguments for spacetimes with spatial boundary by imposing suitable boundary conditions.} I may choose the boundary to be given by $\partial U=\Sigma_{t_1} \cup \Sigma_{t_2}$ where $\Sigma_{t_1}$ is a spacelike constant time hypersurface defined by the coordinate value $t=t_1$, and $\Sigma_{t_2}$ is defined by $t=t_2$. Equation (\ref{CCGRA-ConservedCurrentDivInt}) then implies the following:
\begin{equation} \label{CCGRA-ConservedIntegralKV}
\int_{\Sigma_{t_2}} n_\mu \, J^\mu \, \sqrt{|h|} d^3y = \int_{\Sigma_{t_1}} n_\mu \, J^\mu \, \sqrt{|h|} d^3y .
\end{equation}

\noindent Since the above holds for an arbitrary coordinate system on the spacetime manifold, one may infer that the following integral is conserved (assuming spacelike $\Sigma$), where the factor $1/4\pi$ is chosen so to yield the appropriate value of mass for an asymptotically flat spacetime:
\begin{equation} \label{CCGRA-ConservedQuantityKV}
Q = \frac{1}{4 \pi} \int_{\Sigma} n_\mu \, J^\mu \, \sqrt{|h|} d^3y .
\end{equation}

\noindent In particular, Eq. (\ref{CCGRA-ConservedIntegralKV}) implies that $Q$ is independent of the choice of spacelike hypersurface $\Sigma$ in the spacetime; it is a conserved quantity.



\bibliography{bibcurrents}

\end{document}